\documentclass[12pt]{article}
\usepackage{subcaption}
\usepackage{graphicx}
\usepackage{multicol}
\usepackage{color}
\usepackage{caption}
\usepackage{multirow}
\usepackage{amsthm}

\usepackage{setspace}
\usepackage{url}
\usepackage{amssymb,amsmath}
\usepackage{amsfonts}
\usepackage{bm}          
\usepackage{tabularx}
\usepackage{enumerate}
\usepackage{comment}
\usepackage{longtable}
\usepackage{algorithmic}
\usepackage[nottoc,notlot,notlof]{tocbibind}

\usepackage{natbib}
\usepackage{mathrsfs}
\usepackage{lscape}

\newcommand{\blind}{1}
\newcommand{\Pb}{\mathbb{P}}
\newcommand*{\N}{\mathbb N}
\newcommand*{\HH}{\mathcal{H}}
\newcommand*{\E}{\mathbb E}
\usepackage{lineno}
\usepackage[inline]{trackchanges} 

\begin{document}

\def\spacingset#1{\renewcommand{\baselinestretch}%
{#1}\small\normalsize} \spacingset{1}


\if1\blind
{
  \title{\bf A Multidimensional Fractional Hawkes Process for Multiple Earthquake Mainshock Aftershock Sequences}
  \author{Louis Davis$^{1,2}$\hspace{.2cm},    
   Boris Baeumer$^1$,    
    Ting Wang$^1$\\
    $^1$Department of Mathematics and Statistics, University of Otago,\\
    P.O. Box 56, Dunedin 9054, New Zealand\\
    $^2$E-mail: \url{davislrs2000@gmail.com}}

  \maketitle
} \fi

\if0\blind
{
  \bigskip
  \bigskip
  \bigskip
  \begin{center}
    \LARGE\bf A Multidimensional Fractional Hawkes Process for Earthquakes
\end{center}
  \medskip
} \fi

\begin{abstract}
Most point process models for earthquakes currently in the literature assume the magnitude distribution is i.i.d. potentially hindering the ability of the model to describe the main features of data sets containing multiple earthquake mainshock aftershock sequences in succession. This study presents a novel multidimensional fractional Hawkes process model designed to capture magnitude dependent triggering behaviour by incorporating history dependence into the magnitude distribution. This is done by discretising the magnitude range into disjoint intervals and modelling events with magnitude in these ranges as the subprocesses of a mutually exciting Hawkes process using the Mittag-Leffler density as the kernel function. We demonstrate this model's use by applying it to two data sets, Japan and the Middle America Trench, both containing multiple mainshock aftershock sequences and compare it to the existing ETAS model by using information criteria, residual diagnostics and retrospective prediction performance. We find that for both data sets all metrics indicate that the multidimensional fractional Hawkes process performs favourably against the ETAS model. Furthermore, using the multidimensional fractional Hawkes process we are able to infer characteristics of the data sets that are consistent with results currently in the literature and that cannot be found by using the ETAS model.

\end{abstract}

\noindent%

\spacingset{1.45} 
\section{Introduction}
Self exciting point processes have been used to model earthquake catalogues for the past half century with the first such example being \cite{hawkes1973cluster}. Since then, the \textit{Epidemic type aftershock sequence} (ETAS) model first studied by \cite{Ogata1988} has been used widely and since extended to account for the spatial location of events \citep[e.g.][]{Ogata1998,Ogata2006} or for non-stationary earthquake behaviour caused by stress changes from non-seismic sources \citep[e.g.][]{kumazawa2014nonstationary}. In its most general form it is a mark separable Hawkes process models, with i.i.d. marks, and has the conditional intensity function 

\begin{multline}\label{eq:generalHP}
\lambda(t,M|\HH_t)=\lim_{h,\Delta \to 0}\frac{\E\left[N\left([t,t+h)\times [M,M+\Delta) \right)|\HH_t \right]}{h\Delta}\\=f(M)\lambda(t|\HH_t)=f(M)\left(\mu+ \sum_{i:t_i<t}g(t-t_i,M_i) \right)
\end{multline}
 where $N(X)$ is the number of points in the Borel set $X$, $\HH_t$ is the history of the process formally defined as the set of events $\{(t_i,M_i):t_i<t\}$, $f(M)$ is the mark density, $\mu$ is the Poisson immigrant rate and $g(t,M)$ is the kernel function, which is integrable on $[0,\infty)$ with respect to $t$. For the ETAS model, $g(t,M)$, as well as $f(M)$, in Equation \eqref{eq:generalHP} have empirically derived forms. Often $g(t,M)=\exp\{\delta(M-M_0)\}\frac{A}{(t/c_E+1)^p}$, where $M_0$ is the minimum magnitude of the catalogue such that all events above this magnitude are detected, and $A,\delta,c_E,p$ are parameters. $g(t,M)$ is derived from two empirical laws, the Omori-Utsu law \citep{utsu1961statistical} saying that the number of aftershocks of an event decays as a power law, and the Utsu law \citep{utsu1971aftershocks} that says the number of aftershocks grows exponentially with the triggering event's magnitude. Furthermore, the marks (magnitudes of earthquakes) are exponentially distributed agreeing with the Gutenberg-Richter law \citep{Gutenberg1944} meaning that $f(M)$ is often parameterised as $b\exp\{-b(M-M_0)\}$. Despite these complex triggering mechanisms, the immigrant rate is simple, the magnitude distribution is independent of the history, and the model as a whole is analytically intractable, hence the long range forecasting ability of the ETAS model is limited \citep[e.g.][]{harte2013bias}. This is further supported by \cite{zhang2020scaling} who suggest the role of memory is weaker (stronger) in the short (long) time scale for the ETAS model than in actual seismic data and \cite{harte2019evaluation} who found that the ETAS model under predicts the number of aftershocks immediately following a main event and over predicts this number later into the aftershock sequence. All of these studies indicate the ETAS model may not be suitable for earthquake forecasting over most time scales of interest. 

Other models in the literature incorporate history dependence into the arrival time distribution of immigrant events \citep[e.g.][]{chen2018direct,STINDL2018131,kolev2019inference}. Most often, history dependence of immigrant events is achieved by using a renewal background intensity. For example $\lambda(t|\HH_t)=\lambda_0(t|\HH_t)+\sum_{i:t_i<t}g(t-t_i,M_i)$
where $\lambda_0(t|\HH_t)=\frac{f\left(t-t_n\right)}{1-F(t-t_n)}$, $t > t_n$ for $t_n$ being the time of the most recent immigrant arrival and $F(t)=\int_{-\infty}^t f(x)dx$ is a CDF with density $f$. This is a relatively simple way to incorporate history dependence into the arrival time of immigrant events. However, parameter estimation of a Hawkes process with a renewal immigrant rate is nontrivial since it is not known which events are immigrants. \cite{WHEATLEY2016120} developed an EM algorithm for parameter estimation, however the computation time of their algorithm scales poorly as the size of the data set increases. \cite{chen2018direct,STINDL2018131} developed a likelihood based method to estimate parameters, and \cite{stindl2023algorithm} modified the algorithm of \cite{WHEATLEY2016120}, all to avoid the aforementioned poor computational scaling making parameter estimation feasible for large data sets. 
 
Other models in the literature that describe long term earthquake behaviour are Markov modulated point processes \citep[e.g.][]{Wang2012} or Hidden Markov Models \citep[e.g.][]{wang2017hidden,orfanogiannaki2018multivariate}. These are models that have an underlying memoryless Markov chain with a finite number of discrete states governing their behaviour. Overall, these modelling assumptions may fail to incorporate the high order interactions that are seen between events and their magnitudes \citep[e.g.][]{helmstetter2003foreshocks,bebbington2010repeated,zhang2020scaling}. However, like the ETAS model, none of the aforementioned models aim to incorporate history dependence into the magnitude distribution.

\cite{chen2021fractional} and \cite{https://doi.org/10.48550/arxiv.2211.02583} studied a fractional Hawkes process which is a Hawkes process model with kernel function being a Mittag-Leffler density. The fractional Hawkes process is similar to the ETAS model since its kernel function asymptotically decays as a power law. Furthermore, because of the analytic properties of the Mittag-Leffler function \cite{chen2021fractional} were able to easily derive closed form values for quantities of interest, such as the average expected intensity.  \cite{davis2024fractional} altered the model of \cite{chen2021fractional} and \cite{https://doi.org/10.48550/arxiv.2211.02583} so that it reflected the empirical law that larger magnitude earthquakes produce exponentially more aftershocks \citep{utsu1971aftershocks}, and incorporated a time scaling parameter to connect the model to the fractional Zener model of \cite{metzler2003fractional}. Explicitly, it was a mark separable Hawkes process model with conditional intensity function, as in Equation \eqref{eq:generalHP}, such that $g(t-t_i,M_i)=\alpha \exp\{\gamma(M_i-M_0)\}cf_\beta(c(t-t_i))$, where $cf_\beta(ct)$ is the time scaled Mittag-Leffler density, $cf_\beta(ct):=c^\beta t^{\beta-1}E_{\beta,\beta}(-(ct)^\beta)$, and $E_{\beta,\beta}(z)$ is the two variable Mittag-Leffler function defined by the power series $E_{a,b}(z):=\sum_{n=0}^\infty \frac{z^n}{\Gamma(na+b)}$ for $a,b \in (0,1]$.

 \cite{davis2024fractional} found that on data sets with relatively stronger clustering of aftershocks their model more suitably described the main features of the data than the ETAS model. However, for the same reasons as the ETAS model its long term forecasting ability is likely also limited. 

The objective of this study is to construct a multidimensional Hawkes process model that can be used to model both history and size dependent activities for point processes with long term records, such as an earthquake catalogue consisting of several mainshock aftershock sequences with the eventual goal being to produce better forecasts than is currently possible. We introduce and motivate a new model in Section \ref{sec:MDFHP}, which uses the same kernel function as \cite{davis2024fractional} due to its desirable analytic properties. In Section \ref{sec:MDFHPdata&resid} we fit both the ETAS and multidimensional fractional Hawkes process model to two long term data sets, one being Japan and the other the Middle American Trench both containing multiple mainshock aftershock sequences. We then compare the information criteria for both models and perform residual analysis. Finally, in Section \ref{subsec:ch6predperformance}, we investigate the information gain of the multidimensional fractional Hawkes process model and once again compare it to the ETAS model. 

\section{The Multidimensional Fractional Hawkes Process (MDFHP)}\label{sec:MDFHP}

The multidimensional Hawkes process is a generalisation of the Hawkes process to a vector valued process first studied by \cite{Hawkes1971}. Its conditional intensity function is a vector valued stochastic process $\bm \lambda(t|\HH_t)=(\lambda_1(t|\HH_t),\dots \lambda_{N_b}(t|\HH_t))$, such that $\forall i \in \{1,2,\dots {N_b}\}$, each component is of the form $\lambda_i(t|\HH_t)=\lambda_{0i}+\sum_{j=1}^{N_b}\sum_{\ell:t_{\ell,j}<t}g_{ij}(t-t_{\ell,j})$ and has the interpretation that it is the conditional intensity function for the $i^{th}$ subprocess. Furthermore, $g_{ij}$ is a nonnegative integrable function on $[0,\infty)$, $\lambda_{0i}$ is the Poisson immigrant rate of events in subprocess $i$ and $t_{\ell,j}$ is the $\ell^{th}$ arrival in subprocess $j$.

We construct the desired dependence between the event size and history in an earthquake catalogue by first discretising the allowable magnitudes into a number of disjoint intervals with finite length. The minimum magnitude, $M_0$, is selected as the lowest magnitude where the Gutenberg-Richter law holds. The maximum magnitude is set at 10, the upper limit for any terrestrial earthquake \citep{USGSmaxmag}. We then discretise the magnitude range as $[M_0,M_m]=\bigcup_{i=1}^{N_b}[M'_{N_b-i-1},M'_{N_b-i}).$ 

We then set each subprocess to correspond to the events in one of the disjoint intervals so that the superposition of these subprocesses describes the entire process. Furthermore, the marks of each subprocess will be modelled by an i.i.d. truncated exponential distribution. For the $i^{th}$ subprocess the density of this distribution is
\begin{equation}\label{eq:markdensity}
f_i(M)=\begin{cases}
\frac{B_{ii}\exp\left\{-B_{ii}\left(M-M'_{N_b-i}\right)\right\}}{1-\exp\left\{-B_{ii}\left(M'_{N_b-i+1}-M'_{N_b-i}\right)\right\}}, & M \in [M'_{N_b-i},M'_{N_b-i+1})\\
0, & \text{ otherwise.}
\end{cases}\end{equation}
 Therefore, the conditional intensity function of the $i^{th}$ subprocess of the multidimensional fractional Hawkes process (MDFHP) is $\lambda_i(t|\HH_t)f_i(M)$ where the ground intensity is
\begin{equation}\label{eq:MDFHPlambi}
    \lambda_i(t|\HH_t)=\lambda_{0i}+\sum_{j=1}^{N_b}\alpha_{ij}\sum_{\ell:t_{\ell,j}<t}\exp\{\gamma_{ij}(M_{\ell,j}-M_0)\}c_{ij}f_{\beta_{ij}}(c_{ij}(t-t_{\ell,j})).
\end{equation}

Note the assumption of each subprocess having an i.i.d. mark distribution still allows for an event's magnitude to depend on the history. Consider the $l^{th}$ event arriving at time $t_l$ with magnitude $M_l$. Then adapting Example 7.5(c) of \cite{daley2003introduction} we have 
\begin{equation}\label{eq:ProbSubproc}
    \Pb[M_l \in [M'_{N_b-i-1},M'_{N_b-i})]=\frac{\lambda_i(t_l|\HH_{t_l})}{\sum_j \lambda_j(t_l|\HH_{t_l})}.
\end{equation}
Equation \eqref{eq:ProbSubproc} says that the possible support of $M_l$ depends on the entire history of the process, and hence the magnitude distribution of the MDFHP model is history dependent despite the i.i.d. mark distribution for each subprocess.

Moreover, constructing the subprocesses this way explicitly allows earthquakes of different sizes to affect the overall process in a magnitude dependent manner. Therefore, we have parameterised the complex dependencies between earthquakes of differing sizes and so by examining the parameter values, and their associated asymptotic confidence intervals, we may be able to discern these relationships. For example, $\lambda_{0i}$ is the immigrant rate of the $i^{th}$ subprocess. The diagonal terms in Equation \eqref{eq:MDFHPlambi}, $\alpha_{ii}\exp\{\gamma_{ii}(M_{\ell,i}-M_0)\}c_{ii}f_{\beta_{ii}}(c_{ii}(t-t_{\ell,i}))$,
describe how the $\ell^{th}$ event in subprocess $i$ excites the conditional intensity of that same subprocess, in an analogous way to the univariate case and hence these are the ``self exciting" terms. The cross terms in Equation \eqref{eq:MDFHPlambi}, when $i \neq j$, $\alpha_{ij}\exp\{\gamma_{ij}(M_{\ell,j}-M_0)\}c_{ij}f_{\beta_{ij}}(c_{ij}(t-t_{\ell,j}))$, describe how the $\ell^{th}$ event in the $j^{th}$ subprocess excites the $i^{th}$ subprocess, and hence are known as the ``mutually exciting" terms.

As far as we are aware thinning the entire earthquake process, by magnitude, into interacting subprocesses has never been done before. However, multidimensional models have been applied to earthquake catalogues previously by thinning the catalogue in the spatial dimension. For example, \cite{liu1998principle,shi1998application} introduced a modification of the stress release model which allowed the $n$ disjoint subregions of the study zone to affect stress changes in other subregions differently.   

In a similar manner, \cite{STINDL2018131} modelled earthquake occurrences using a multivariate Hawkes process where the different subprocesses corresponded to earthquakes occurring in different subregions of the data. Specifically, earthquakes were designated to subprocess 1 if they occurred in Fiji, and subprocess 2 if they occurred in Vanuatu. Their model showed promise since they were able to forecast the total number of events, and the number of those that occurred in the two spatial subregions. This provides hope that our model may be able to successfully forecast the magnitude ranges of future events.

\section{Data Selection, Parameter Estimation and Residual Analysis}\label{sec:MDFHPdata&resid}

\begin{figure}[h!]
    \centering    \includegraphics[width=0.8\textwidth]{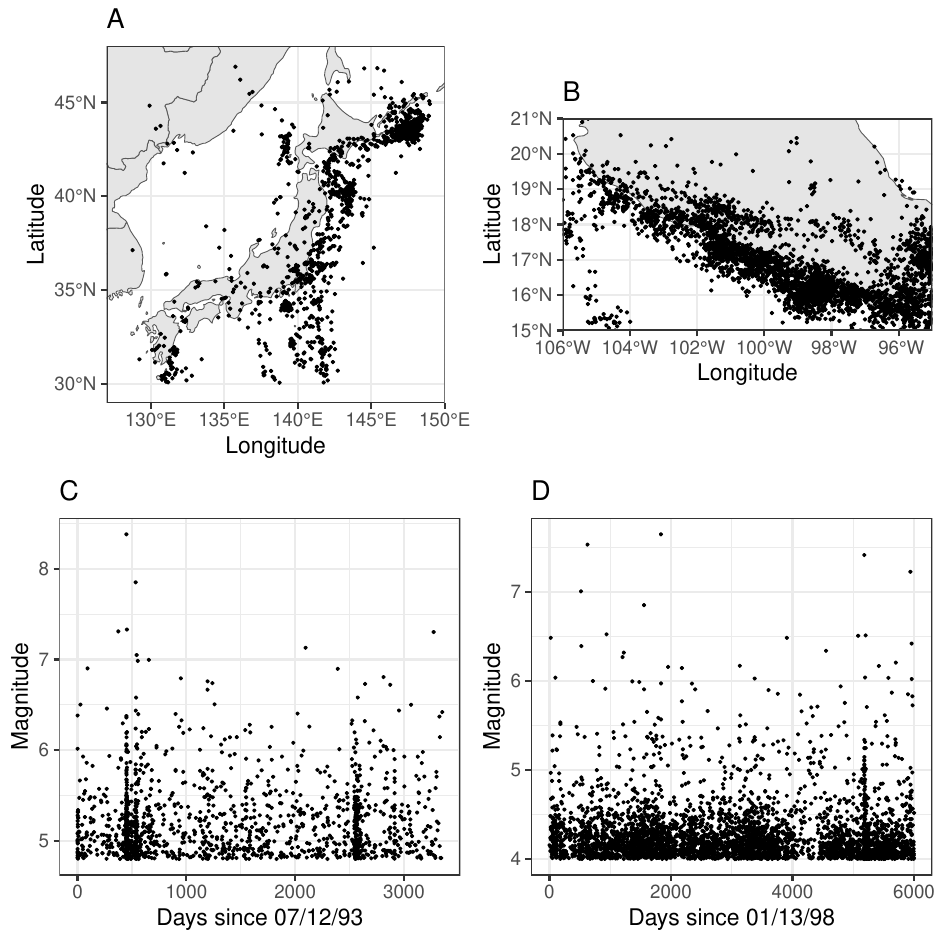}
    \caption{Occurrence maps and magnitude versus time plots for the data sets (A) \& (C) Japan and (B) \& (D) Middle America Trench. }
    \label{fig:MAPS}
\end{figure}

We study two different data sets with multiple mainshock aftershocks sequences, one being all of Japan over a 9 year period from the rectangular spatial region $[128^\circ \text{E},149^\circ \text{E}]\times [30^\circ \text{N},47^\circ \text{N}]$, and the other the Middle America Trench over a period of 16 years in the rectangular spatial region $[106^\circ \text{W},95 ^\circ \text{W}]\times [15^\circ \text{N},21^\circ \text{N}]$ last accessed from the USGS Earthquake Catalogue, on 04/20/22 for the Japan catalogue and 02/28/24 for the Middle America Trench Data Set. We select the Japan data set because this area is one of the most seismically active regions in the globe. Additionally, its time window was chosen so that two periods of high seismic activity were seen with a relatively quiescent background period before and after each. We also study the Middle America Trench for a comparison to the Japan data. The Middle America trench is also a seismically active region but contains only one boundary between tectonic plates \citep[e.g.][Figure 1]{ramirez2020sand}, whereas the study zone  of the Japan data set is a superposition of rebound plates \citep{wakita2013geology}. The temporal duration of the Middle America Trench data set was selected to be when the completeness magnitude was $4$ for the majority of the time window. Summary information for these data sets can be found in Table \ref{tab:Geographicinfo}, as well as an occurrence map and magnitude versus time plots in Figure \ref{fig:MAPS}.

\begin{table}[h]

    \centering
    
    \begin{tabular}{|c c c c c |}
    
    \hline
Catalogue & $N(T)$ & $M_0$ & Time of first event&Time of last event   \\   
    Japan  &  1501 & 4.75 &07/12/93 14:27:20 &09/15/2002 08:39:32 \\ 
    Middle America Trench & 4135 & 4 & 01/13/98 08:30:36 & 06/19/14 00:08:30 \\
    
      \hline 
    \end{tabular}   

\caption{ Summary information of the data sets, where $N(T)$ is the size of the data set, $M_0$ the minimum magnitude for completeness as determined by the GR law and times are in UTC.}
\label{tab:Geographicinfo}
\end{table}

We estimate parameters via maximum likelihood for both the multidimensional Hawkes process and ETAS model. For a multidimensional marked point process the log-likelihood is
\begin{equation}\label{eq:MDloglik}
    \ell(\bm \theta)=\sum_{j=1}^{N_b}\left\{ \sum_{\substack{k \in K_{N_b-j+1}}}\left[ \log(\lambda_j(t_k|\HH_{t_k}))+\log(f_j(M_k))\right]-\int_0^T \lambda_j(t|\HH_t)dt\right\},
\end{equation}
for data $\{(t_i,M_i)\}_{1\leq i \leq N(T)} \subset [0,T] \times [M_0,10]$ with $K_{N_b-j+1}$ being the set of events for the $j^{th}$ subprocess \citep[e.g.][]{bowsher2007modelling}. Equation \eqref{eq:MDloglik} is the sum of the log-likelihoods for each of the subprocesses. 

When maximising Equation \eqref{eq:MDloglik} for the MDFHP model the Mittag-Leffler function was computed using the ``ml.m" function obtained from the MATLAB Central File Exchange by \cite{Garrappa2022A}. This function uses a sophisticated optimal parabolic contour algorithm developed by \cite{garrappa2015numerical} which had parts of it translated into C++ for more efficient computation. Furthermore, for better computational efficiency we also implemented power series and Poincar\'e asymptotic approximations to the Mittag-Leffler function \citep[e.g.][]{haubold2009mittagleffler}. 

While asymptotic properties for stationary, univariate Hawkes process models have been known for over four decades \citep[e.g.][]{ogata1978asymptotic}, the asymptotic properties of maximum likelihood estimates for multidimensional processes have not yet been established analytically \citep[e.g.][]{yang2018applications}. However, \cite{bowsher2007modelling} provides numerical evidence that the maximum likelihood estimates transformed to the log space are approximately normally distributed. Hence, we will assume that asymptotic normality has been reached in the transformed parameter space for the following model analysis. 

We compare the multidimensional model to the ETAS model with an i.i.d. truncated exponential distribution with density $f(M)=\frac{B_E\exp\left\{-B_E(M-M_0)\right\}}{1-\exp\left\{-B_E(10-M_0)\right\}},$ $M \in [M_0,10]$.

We use the CRAN package ``PtProcess" \cite{Harte2010} to fit the parameters in the ETAS model, and a MATLAB script to estimate $B_E$ via maximum likelihood. We can do so since the log-likelihood is separable. 

We select MDFHP models with two subprocesses to keep the parameter space relatively low dimensional. For the Japan data set we discretise the magnitude range as $[4.75,5.5),[5.5,10]$ and the Middle America Trench data set as $[4,4.35),[4.35,10)$. These discretisations were selected qualitatively by choosing a magnitude that earthquakes of at least this size occurred less frequently during periods of relative quiescence. Additionally, each subprocess contains no fewer than 20\% of the data set which means both the subprocesses regularly interact. Developing objective methods to select the subprocesses is a topic of future research. A sensitivity analysis of the choice of the cutoff magnitude for the subprocesses is presented in Appendix \ref{app:subprocess} which suggests a rule of thumb for selecting the subprocesses. 

The maximum likelihood estimates for the ETAS and MDFHP models are presented in Tables \ref{tab:mleMDFHP6} and \ref{tab:CH7ETAS}. We compare the MDFHP and ETAS models by using both the Akaike Information Criterion, $\mbox{AIC}=-2\ell(\hat{\bm \theta}|\bm x)+2k$ \citep{akaike1974new} and Bayesian Information Criteria $\mbox{BIC}=-2\ell(\hat{\bm \theta}|\bm x)+k\log(N(T))$ \citep{schwarz1978estimating}, where $N(T)$ is the size of the data set and $k$ the number of parameters. The best model is the one with the smallest AIC and BIC since this model most effectively balances over parameterisation and goodness of fit. We compare both criteria because the MDFHP has many more parameters than the ETAS model. Hence, the larger penalty term in the BIC may favour selection of the more parsimonious ETAS model. The AIC and BIC values presented in Table \ref{tab:AICBICMDFHP} provide strong evidence that the MDFHP is the best performing model for both data sets. We explore possible reasons why in the following subsection.

\begin{table}[h!]
        \centering
        \begin{tabular}{|ccccccccccc|}
        \hline
        && \multicolumn{4}{c}{Japan (Index)} && \multicolumn{4}{c|}{Middle America Trench (Index)} \\
          Parameter && 11 & 12 & 21 & 22 && 11 & 12 & 21 & 22\\          
          $\hat\lambda_0$ && 0.029 & NA & NA & 0.097 && 0.078 & NA & NA & 0.049\\ 
          $\hat\alpha$  && 0.007 & 0.112 & 0.005 & 0.472& & 0.041 & 0.042 & 0.116 & 0.808 \\ 
          $\hat\gamma$ && 2.142& 0.119 & 2.762 & 0.959 && 1.333 & 3.583 & 1.207 & 0.392\\ 
          $ \hat\beta $ && 0.759 & 0.425 & 0.868 & 0.531 && 0.718 & 0.668 & 0.687 & 0.623\\ 
          $\hat c$ && 5.452 & 0.033 & 2.497 & 0.152 && 11.469 & 0.462 & 2.583 & 0.065\\ 
          $\hat B$ && 2.636 & NA & NA & 2.666 && 2.469 & NA & NA & 7.839 \\\hline
        \end{tabular}
        \caption{Maximum likelihood estimates for the multidimensional fractional Hawkes process models fit to data sets.}
        \label{tab:mleMDFHP6}
\end{table}

\begin{table}[h!]
    \centering
    \begin{tabular}{|lcccccc|}
    \hline
        Catalogue &$\hat\mu$& $\hat A$ & $\hat\delta$ & $\hat c_E$ & $\hat p$ & $\hat B_E$  \\
       Japan & 0.120 & 1.246 & 1.597 & 0.029 & 1.089 & 2.410\\ 
       Middle American Trench & 0.119 & 1.767 & 1.135& 0.022&0.962 & 4.280\\\hline
    \end{tabular}
    \caption{Maximum likelihood estimates for the ETAS model fit to the Japan and Middle America Trench data sets.}
    \label{tab:CH7ETAS}
\end{table}

\begin{table}[h!]
        \centering
        \begin{tabular}{|cccc|}
        \hline
          Catalogue &Model &  AIC & BIC \\
         
          \multirow{2}*{Japan} &ETAS    & 3229.0 & 3260.9\\ 
          &MDFHP  & 3146.0 & 3252.3\\

         \multirow{2}*{Middle America Trench}& ETAS  & 5784.7 & 5822.7 \\
          &MDFHP & 4967.8 & 5094.4\\\hline 
        \end{tabular}
        \caption{AIC and BIC scores for the ETAS and MDFHP models.}
        \label{tab:AICBICMDFHP}
\end{table}

\subsection{Residual Analysis and Statistical Inference}\label{subsec:ch6ResidAnalysis}

We perform residual analysis using the transformed time method given in  \cite{bowsher2007modelling} and Proposition 14.6.V of \cite{daley2008introduction} which is outlined as follows. Suppose that $t_{k,j}$ is the $k^{th}$ event in the $j^{th}$ subprocess. Then the process $\left\{\tau_k^{(j)} \right\}_{1\leq k \leq N_j(T)}$ where $\tau_k^{(j)} =\int_0^{t_{k,j}}\hat \lambda_j(t|\HH_t)dt$ is a Poisson process of unit rate if $\hat \lambda_j(t|\HH_t)$ suitably approximates the conditional intensity function of the $j^{th}$ data generating subprocess. Moreover, $\left\{\tau_k^{(i)} \right\}_{1\leq k \leq N_i(T)}$ is independent of $\left\{\tau_k^{(j)} \right\}_{1\leq k \leq N_j(T)}$ when $i \neq j$. We analyse the residual process in three ways. We first plot the mean removed residual process against the event number, which should be approximately constant around $0$ if the model fits the data well. Furthermore, since $\left\{\tau_k^{(j)} \right\}_{1\leq k \leq N_j(T)}$ should be arrivals of a Poisson process of unit rate we test if the transformed uniform inter-event times, $U_k^{(j)}=1-\exp\left\{-\left(\tau_{k}^{(j)}-\tau_{k-1}^{(j)}\right)\right\},$ are i.i.d. uniformly distributed random variables on $[0,1]$ by conducting a one sided Kolmogorov-Smirnov (KS) test and Pearson correlation test of $U_k^{(j)}$ against $U_{k+1}^{(j)}$.

\begin{figure}[h!]
    \centering
    \includegraphics[width=\textwidth,trim={0 0 0 0},clip]{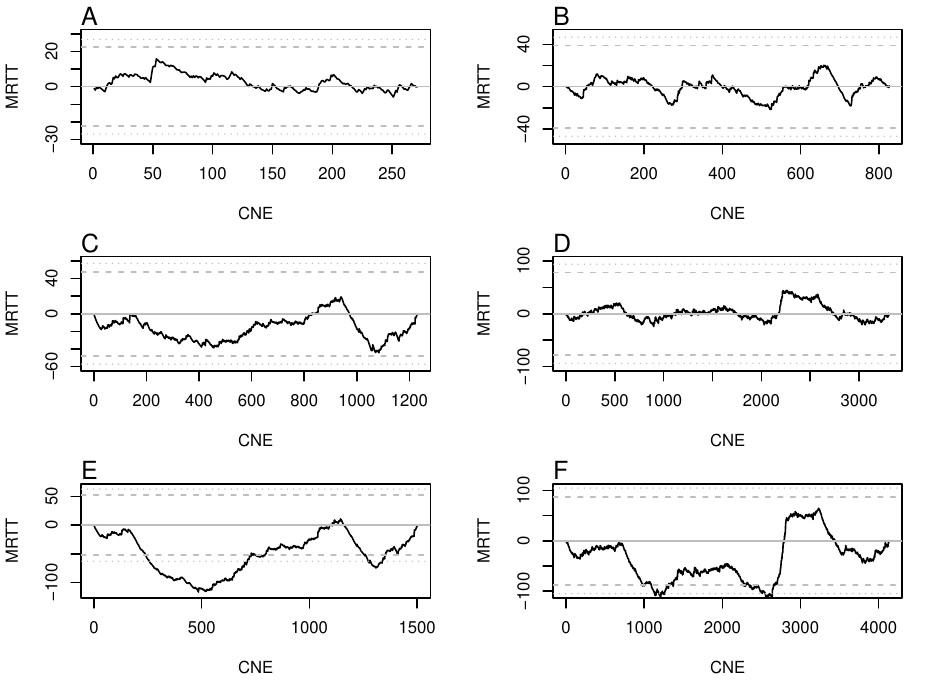}
    \caption{Mean removed transformed time residuals (MRTT) against the cumulative number of events (CNE). For all plots the black curve is the mean removed residual process, the solid gray line is the theoretical mean (0) and the dashed and dotted grey lines are the 95\% KS and 99\% KS confidence intervals respectively. A,C,E correspond to subprocess 1, 2 and the ETAS model respectively fit to the Japan data set. B,D,F correspond to subprocess 1, 2 and the ETAS model respectively fit to the Middle America Trench data set. }
    \label{fig:JASAComp}
\end{figure}

Figure \ref{fig:JASAComp} plots the mean removed transformed time process with 95\% and 99\% KS confidence intervals for the ETAS and MDFHP models respectively. We see that the ETAS model greatly exceeds the lower 99\% confidence bound for both data sets. However, both subprocesses of each multidimensional fractional Hawkes process stay within the 95\% KS confidence bound. Furthermore, the results of the Pearson and KS tests in Table \ref{tab:PearsonPvaluesMDFHP} suggests for both data sets we do not have evidence to reject the null hypothesis that the transformed uniform inter-event times of the MDFHP  are uniformly distributed for every subprocess. However, we have evidence to reject the null hypothesis that $U_k$ and $U_{k+1}$, corresponding to the ETAS model fitted to the Japan data set, are uncorrelated. To the contrary, at the 95\% significance level we do not have sufficient evidence to reject the null hypothesis that $U_k$ and $U_{k+1}$ corresponding to the ETAS model fitted to the Middle America Trench data are uncorrelated.

\begin{table}

\centering
\begin{tabular}{|cccccc|}
\hline
& & \multicolumn{2}{c}{Pearson Correlation} & \multicolumn{2}{c|}{KS test}\\
Data Set &Model& Test statistic& $p$ value & Test statistic & $p$ value\\
 & & & & &  \\
\multirow{3}*{Japan} & ETAS  &4.913&$<10^{-6}$ & 0.023 & 0.400\\
 & MDFHP SP1 &-0.077&0.937&0.040 & 0.785\\
  &MDFHP SP2 &1.699&0.090&0.025& 0.394\\

 & & & & &  \\
\multirow{3}*{Middle America Trench}&ETAS & 1.960 & 0.050 & 0.019 & 0.103\\
&MDFHP SP1 & 0.935 & 0.350 & 0.033 & 0.318\\
&MDFHP SP2 & 1.044 & 0.295 & 0.013 & 0.611\\
\hline
\end{tabular}
\caption{ Test statistics and $p$ values of the Pearson test for serial correlation between $U_k$ and $U_{k+1}$ and KS test for uniformity of $U_k$ where ``SP1" and ``SP2" stand for ``subprocess 1" and ``subprocess 2" respectively.}
\label{tab:PearsonPvaluesMDFHP}
\end{table}

Overall, we have strong evidence to suggest that the MDFHP is describing the main features of both data sets sufficiently. However, the qualitative residual analysis test suggests some features of the data are unexplained by the ETAS model. The discrepancies between the ETAS model and data, especially on the Japan data set, can be explained by examining the maximum likelihood estimates of both models and the 90\% asymptotic confidence intervals for the MDFHP5.5 model.

We first analyse the differences between the ETAS and MDFHP model on the Japan data set. The residual process corresponding to the ETAS model, Figure \ref{fig:JASAComp} (E), exceeds the 99\% confidence intervals in the aftermath of the two distinct main shocks in the catalogue (events 158 and 1142). The reason for this is that too few events are expected from the fitted model. This indicates that the ETAS model is unable to switch between the relatively low seismic activity rate outside of the aftershock period and the relatively high rate inside of it (see Figure \ref{fig:JASAComp} (E) around transformed time 500 and just after 1250). There are relatively more large earthquakes ($M\geq 5.5$) around the two major clusters than in the background periods (Figure \ref{fig:MAPS} (C)). The variable rate productivity, $\exp\{\hat \delta(M_i-M_0)\}$, is then of the utmost importance since this will be a major contributing factor to the model being able to increase the seismicity rate in response to large earthquakes. For the ETAS model, $\hat \delta=1.597$ meaning for every event the intensity is increased by a term proportional to $\exp\{1.597(M-M_0)\}$. Compared to the MDFHP models this causes a smaller increase in the intensity of the ETAS model when events $M\geq 5.5$ arrive, which is clear by observing the values of $\hat\gamma_{11}$ and $\hat\gamma_{21}$, which are bigger than $\hat \delta$, in Table \ref{tab:mleMDFHP6}. Therefore, these larger events increase the observed intensity of the MDFHP models more than in the ETAS model, and since they occur more often in times of high seismicity, the MDFHP is able to respond more appropriately to the aftershock regime.

Similarly, for the Middle America Trench data set the MDFHP is more flexible, hence its response to the time varying behaviour of the data is more appropriate. The residual process of the ETAS model breaches the lower confidence intervals around events 1000 and 2500, which are periods of greater clustering of small events around 1500 and 3500 days after the start of the catalogue, as seen in Figure \ref{fig:MAPS} (D). The reason that the ETAS model exceeds the lower bounds here is that at time 4000 (approximately event 2700) a period of extremely low seismic activity begins and continues for approximately 700 days. This time period is influencing the ETAS model so that it does not react as much to any given event, as otherwise it would over predict the number of events during this relatively quiescent period. As a result, during the aforementioned periods of greater seismic activity the conditional intensity of the ETAS model does not increase enough and so under predicts the number of events. On the other hand, each subprocess of the MDFHP has different triggering characteristics for each subprocess. Therefore it can more easily adapt to the changing seismic behaviour, which is best highlighted by computing the expected number of offspring for an event. Following the logic of \cite{harte2013bias} Subsection 2.2, the expected number of offspring in subprocess $j$, triggered by an event in subprocess $i$ in the first generation, is
\begin{equation}
    \E[\alpha_{ji}\exp\left\{\gamma_{ji}(M_{l,i}-M_0)\right\}]= \alpha_{ji}\int_{0}^{M_{N_b-i+1}-M_{N_b-i}} \frac{\exp\{(\gamma_{ji}-B_{ii})M\}}{1-\exp\{-B_{ii}(M_{N_b-i+1}-M_{N_b-i})\}}dM
\end{equation}
because the kernel function is a density in time (hence its integral over $[0,\infty)$ is $1$). An event of magnitude $M < 4.35$ is expected to trigger 0.1074 events of $M < 4.35$ and only 0.0082 of $M \geq 4.35$. Events $M \geq 4.35$ are expected to trigger 0.036 events in the same magnitude range and 0.0918 of $M<4.35$. Therefore, the magnitude of an offspring is directly dependent on the magnitude of the parent event, which is not the case for the ETAS model due to its i.i.d. magnitude distribution. It is this behaviour that allows the MDFHP to better adapt to the different stages of the seismic cycle since these clusters of small events preferentially trigger other small events which the ETAS model cannot do. Such a result also agrees with \cite{Nichols2014} who found that there is a positive correlation between an earthquake's magnitude and the magnitude of its immediate offspring. 

The AIC and BIC values for the MDFHP models are smaller than that for the ETAS model. This is despite the large penalty term for the MDFHP because of its twenty parameters, as opposed to the ETAS model's six, which means that the log-likelihood of the MDFHP models must be much greater than that of the ETAS model. To examine why, recall the log-likelihood (Equation \eqref{eq:MDloglik}) \[\ell(\bm \theta)=\sum_{j=1}^{N_b} \sum_{\substack{k \in K_{N_b-j+1}}}\left[ \log(\lambda_j(t_k|\HH_{t_k}))+\log(f_j(M_k))\right]-\int_0^T \lambda_j(t|\HH_t)dt.\]
For all of the fitted models $\int_0^T \lambda_j(t|\HH_t)dt= \tau_{N_j(T)}^{(j)} \approx N_j(T)$, and $N_1(T)+N_2(T) =N(T)$. Therefore, the differences in the log-likelihood are entirely due to the sum of the conditional log intensities and magnitude distribution, which is much greater for the MDFHP models than the ETAS model. 

We now examine the sum of the conditional log intensities and magnitude distribution more closely. Consider the two quantities $I:=\sum_{j=1}^{N_b} \sum_{\substack{k \in K_{N_b-j+1}}} \log(\lambda_j(t_k|\HH_{t_k}))$ and $\quad J:=\sum_{j=1}^{N_b} \sum_{\substack{k \in K_{N_b-j+1}}}\log(f_j(M_k))$ for each of the fitted models (with $N_b=1$ for the ETAS model). $I$ is much greater for the ETAS model than for the MDFHP. When considering the Middle America Trench data set, this is likely because the mean intensity of both subprocesses of the MDFHP are less than that of the ETAS model, and since most events occur when $\hat \lambda_i(t|\HH_t)<1$, the mean summand in $I$ is less for the MDFHP compared to the ETAS model. For the Japan data set the mean intensity at the event times of subprocess 2 is greater than the mean intensity at event times for the ETAS model. However, $I$ is still less for the MDFHP model possibly because fewer terms in $I$ are positive (492 for the MDFHP vs 570 for the ETAS model). Furthermore, the mean of $\hat \lambda_1(t_i|\HH_{t_i})$ is less than $\hat \lambda_{\text{ETAS}}(t_i|\HH_{t_i})$ for $t_i$ being an arrival such that $M_i>5.5$. This provides evidence that more terms in $I$ for the MDFHP are smaller than for the ETAS model explaining why $I$ is greater for the ETAS model. However, for both data sets $J$ is greater for the MDFHP likely because the majority of the events are classified to subprocess 2, and their magnitudes are modelled by a random variable with a relatively small support. Therefore, the estimated magnitude density, $\hat f_{2}(M)$, can take a greater value for $M \approx M_0$. This greatly increases the log-likelihood of the MDFHP to the extent that the information criteria favours selection of the MDFHP as the best model.

For the remainder of this subsection we infer characteristics of the data from the MDFHP model fitted to both data sets using the 90\% confidence intervals presented in Table \ref{tab:MDFHPCI}. Starting with the Japan data set we see for the parameter $\hat \alpha_{ij}$ that at the 90\% confidence level $\hat \alpha_{11}<\hat \alpha_{12}$ and $\hat \alpha_{21}<\hat \alpha_{22}$. Furthermore, while the confidence intervals for $\hat \gamma_{11}$ and $\hat \gamma_{12}$, as well as $\hat \gamma_{21}$ and $\hat \gamma_{22}$, overlap, the values in Table \ref{tab:mleMDFHP6} show that  $\hat \gamma_{11}>\hat \gamma_{12}$ and $\hat \gamma_{21}>\hat \gamma_{22}$. Combined, these pieces of information suggest that in both subprocesses the smaller events affect the process in a more uniform manner than the larger events, as in, they cause relatively smaller spikes in the conditional intensity function independent of magnitude. If $\hat \gamma_{ij}$ is smaller, then the expected number of offspring in subprocess $i$, of an event in subprocess $j$, is less dependent on that event's magnitude. Also, if $\hat \alpha_{ij}$ is larger the contribution to the intensity of subprocess $i$, by an event in subprocess $j$, is also greater independent of itsw magnitude.

Furthermore, the values of $\hat \beta_{ij}$ and $\hat c_{ij}$ suggest how the contribution to the intensity of each subprocess decays differently dependent on the size of the event. At the 90\% confidence level $\hat c_{11}>\hat c_{12}$ and $\hat c_{21}>\hat c_{22}$ suggesting that within both subprocesses the influence of larger events on the intensity of each subprocess decays faster than that of the smaller events. Furthermore, at this confidence level, $\hat\beta_{21}>\hat\beta_{22}$ and for the point estimates $\hat \beta_{11}>\hat \beta_{12}$. Asymptotic behaviour of $c_{ij}f_{\beta_{ij}}(c_{ij}t)$ has interesting consequences from the relative sizes of $\hat \beta_{ij}$. As $x\to 0$, $f_{\beta_{ij}}(x)\approx x^{\beta_{ij}-1}$ and as $x \to \infty$, $f_{\beta_{ij}}(x)\approx x^{-\beta_{ij}-1}$. Larger values of $\hat \beta_{ij}$ mean the intensity decays faster both in a neighbourhood of $0$ and also for large values of time. Combined with the estimated values of $\hat c_{ij}$ it is clear the larger events are mostly influential in a neighbourhood of their arrival time, while it is the smaller events that are more likely to have offspring much longer after their arrival time. Since the larger events likely have more offspring than smaller ones and their influence decays faster, we have evidence to suggest that the larger events have more offspring, of any size, closer to their arrival time than the smaller earthquakes do.

We now analyse the Middle America Trench data set. The most obvious feature is that $\hat c_{11}<\hat c_{12}$ and $\hat c_{21}<\hat c_{22}$, implying that the influence of events $M \geq 4.35$ decays faster than events of magnitude $M<4.35$, hence the larger events mostly influence the intensity of each subprocess in a neighbourhood of their event time, as was seen in the Japan data set. Since the expected number of immigrant events of magnitude $M<4.35$ is $\hat \lambda_{02}T=294$ out of 3309, most small events are likely triggered. Furthermore, because $\hat b$ is relatively large, most events in this catalogue are small. These ideas, when used in combination with the fact that the expected number of small offspring triggered from a small event is greater than the number triggered by a larger event, may suggest that events of magnitude $M<4.35$ are more important for the overall triggering behaviour in this data set. Otherwise, there is less we can infer at a 90\% confidence level because most of the remaining asymptotic confidence intervals overlap.

\section{Information gain}\label{subsec:ch6predperformance}
 We finally investigate the information gain per unit time (IGPT) \citep[e.g.][]{vere1998probabilities} of the MDFHP and ETAS models. The IGPT is found by first discretising the catalogue's time domain into two day intervals, with the final interval being the remainder. Not only was this interval length used by \cite{vere1998probabilities} when the IGPT was calculated on a seismic catalogue of a similar duration and number of events, but it is also close to the mean interarrival time of events in the two catalogues (2.2 days for the Japan data set and 1.4 days for the Middle America Trench). Explicitly, our time intervals are $\{(t'_i,t'_{i+1}]:i=1,2,\dots n\}$ where $t'_1=10^{-6}$ which is approximately $0.1$ seconds after the first event. We then simulate $2000$ realisations of each model in the $i^{th}$ time interval using the true history of the process at time $t_i'$ by the thinning method \citep[e.g.][]{lewis1979simulation,ogata1981lewis}. Furthermore, we discretise the magnitude range into $K$ classes $\{[M_k,M_{k+1}):k =0,1,\dots K-1\}$.  $p_i(k)$ is the proportion of simulated trials which contain at least one event in the $i^{th}$ time interval and $k^{th}$ magnitude class. $\bar p_i(k)$ is the probability of observing at least one event in the $i^{th}$ time and $k^{th}$ magnitude interval under the reference process. In our case, the reference process is the empirical rate Poisson process, $\bar \lambda=N(T)/T$, with i.i.d. exponential marks with density $f(M)=b\exp\{-b(M-M_0)\}$ where $b$ is the MLE. Hence, $\bar p_i(k)=\left(1-\exp\{-\bar \lambda(t_{i+1}-t_i)\}\right)\int_{M_k}^{M_{k+1}}f(M)dM.$
 
 Finally, define $X_i(k)$ to be $1$ if there is a data point in the $i^{th}$ time and $k^{th}$ magnitude interval and $0$ otherwise. Then the information gain per event in the $k^{th}$ magnitude class, $\rho_T(k)$, is
\[\rho_T(k)=\frac{1}{T}\sum_i\left\{X_i(k)\log\left[\frac{p_i(k)}{\bar{p_i}(k)}\right]+(1-X_i(k))\log\left[\frac{1-p_i(k)}{1-\Bar{p_i}(k)}  \right] \right\}.\]
Furthermore, we decompose $\rho_T(k)$ into
\[G_S(k)=\sum_{i:X_i(k)=1}\log\left[\frac{p_i(k)}{\bar{p_i}(k)}\right], \quad G_F(k)=\sum_{i:X_i(k)=0}\log\left[\frac{1-p_i(k)}{1-\Bar{p_i}(k)}  \right]\]
and define $G(k)=G_S(k)+G_F(k)$. We can use these statistics, as opposed to those found by summing over $k$, to evaluate the models' information gain for earthquakes of different sizes. 

For the Japan data set we select the magnitude classes as $[4.75,5.5), [5.5,6.5), [6.5,10]$. The smallest magnitude class, $M \in [4.75,5.5)$, was selected to be the same as the smallest subprocess of the MDFHP model since we are not concerned about forecasting smaller earthquakes. We selected the largest class as $M \geq 6.5$ since events of this size have the potential to cause moderate to severe damage, dependent on their location and exact magnitude. The moderate sized events were simply what was left. For the Middle America Trench data set we selected the magnitude classes as $[4,4.35), [4.35,5.35), [5.35,10]$. Once again the smallest subprocess is the same as subprocess 2 of the MDFHP. We selected the largest class as $M \geq 5.35$ since there are only 76 events of at least this magnitude, and if this cutoff magnitude was larger it was more likely that $G_S(2)$ would be undefined (since $p_i(2)=0$ and $X_i(2)=1$). The middle magnitude class is once again what remains. The information gain for the MDFHP and ETAS models are presented in Table \ref{tab:mdfhp55J93IGP}.

\begin{table}[h!]
    \centering
    \begin{tabular}{|cccccccc|}
    \hline
     Mag. class & $N_S(k)$ & $G_S(k)$ &$N_F(k)$ & $G_F(k)$ & $N(k)$ & $G(k)$ & $\rho_T(k)$ \\
      & & & & & & & \\
    \multicolumn{8}{|c|}{MDFHP Japan}\\

        $4.75 \leq M < 5.5$ & 631 & -159.18 & 1045 & 253.30 & 1676 & 94.12 & 0.028\\
        $5.5 \leq M < 6.5$ & 183 & 66.00 & 1493 & -45.13& 1676 & 20.86 & 0.0062\\
        $6.5 \leq M$ &  23 & 8.32 & 1653 & -2.52 & 1676 & 5.79 & 0.002 \\
        & & & & & & & \\

\multicolumn{8}{|c|}{ETAS Japan}\\

        $4.75 \leq M < 5.5$ & 631 & -305.25 & 1045 & 347.02 & 1676 & 41.78 & 0.012\\
        $5.5 \leq M < 6.5$ & 183 & -10.26 & 1493 & 7.96& 1676 & -2.30 & -0.0007\\
        $6.5 \leq M$ &  23 & 6.23 & 1653 & -2.52 & 1676 & 3.71 & 0.001 \\
         & & & & & & & \\

           \multicolumn{8}{|c|}{MDFHP Middle America Trench}\\

        $4 \leq M < 4.35$ & 1705 & 27.49 & 1296 & 124.15 & 3001 & 151.64 & 0.025\\
        $4.35 \leq M < 5.35$ & 569 & 106.86 & 2432 & -90.00& 3001 & 16.86 & 0.003\\
        $5.35 \leq M$ &  71 & 157.53 & 2930 & -56.78 & 3001 & 100.75 & 0.017 \\

 & & & & & & & \\
           \multicolumn{8}{|c|}{ETAS Middle America Trench}\\

        $4 \leq M < 4.35$ & 1705 & -75.30 & 1296 & 183.17 & 3001 & 107.87 & 0.018\\
        $4.35 \leq M < 5.35$ & 569 & 193.99 & 2432 & -211.62& 3001 & -17.63 & -0.003\\
        $5.35 \leq M$ &  71 & 33.59 & 2930 & -5.95 & 3001 & 27.64 & 0.005 \\\hline
    
\end{tabular}
    \caption{Information gain of the ETAS and MDFHP models fitted to both data sets. For the $k^{th}$ magnitude class $N_S(k)$ is the number of intervals such that $X_i(k)=1$ and $G_S(k)$ is the gain in these intervals. $N_F(k)$ is the number of intervals such that $X_i(k)=0$ and $G_F(k)$ is the gain in these intervals. $N(k)$ is the total number of intervals, $G(k)$ is the total information gain, and $\rho_T(k)$ is the information gain per unit time.}
    \label{tab:mdfhp55J93IGP}
\end{table}

For all models, except the MDFHP fit to the Middle America Trench data set, $G_S(0)$ is negative, which means when $X_i(0)=1$ more often than not $p_i(0)<\bar p_i(0)$. This is likely because most small events, of magnitude $M<5.5$ for the Japan data set or $M<4.35$ for the Middle America Trench data set, occur during a period with a relatively low seismic rate. Therefore, these models are scored similar to a Poisson process of rate $\hat \mu$ or $\hat \lambda_{02}$, which are both less than the mean rate due to clustering. However, $G_F(0)$ is positive since generally $1-p_i(0)>1-\bar p_i(0)$ when $X_i(0)=0$ to such an extent $G(0)$ is still positive for all fitted models. The most likely reason that $G_S(0)$ and $G_F(0)$ are both positive for the Middle America Trench data set is that the MDFHP changes in accordance with the periods of there being more small events around time 1500 and 3500, and relative quiescence around time 4250, better than the ETAS model is able to.  

We can observe that there is generally a positive correlation between the estimated conditional intensity of the process, $\hat \lambda(t,M|\HH_t)$, and a more positive value and more negative value of $G_S(k)$ and $G_F(k)$ respectively. As was described previously, the ETAS model has the largest AIC and BIC since $\sum_i \log(\hat \lambda(t_i,M_i|\HH_{t_i}))$ is smaller for the ETAS model, implying that the estimated intensity of the ETAS model is less than that of the MDFHP model which indicates that the ETAS model predicts relatively fewer events than the MDFHP. This explains why $G_S(k)$ is more negative and $G_F(k)$ is more positive for the ETAS model in every magnitude class, except $4.35 \leq M<5.35$ for the Middle America Trench data set, when compared to the MDFHP. 

We now examine this other case by computing the probability $\Pb[M \in [4.35,5.35)]$ given there is an arrival an time $t$. According to the ETAS model this is $\Pb[M \in [4.35,5.35)]=0.224$, which may be found by simple integration of $f(M)$ since the marks are assumed to be i.i.d.. For the MDFHP, we can approximate $\Pb[M \in [4.35,5.35)]$ by averaging 
\begin{equation}\label{eq:MagProb}
    \frac{\lambda_1(t|\HH_t)}{\lambda_1(t|\HH_t)+\lambda_2(t|\HH_t)}\int_{4.35}^{5.35}f_1(M)dM=\frac{\lambda_1(t|\HH_t)\left(1-\exp\{-2.469\} \right)}{\left(\lambda_1(t|\HH_t)+\lambda_2(t|\HH_t)\right)\left(1-\exp\{-13.950\} \right)},
\end{equation}
at event times and $10^{-3}$ days before and after each event. Doing so, we can approximate that $\Pb[M \in [4.35,5.35)] \approx 0.192$. This suggests that the MDFHP expects fewer events in this magnitude range than the ETAS model. This may be because the MDFHP suggests that events of magnitude $M \in [4.35,5.35)$ tend to only be immigrants or triggered by larger rarer events, since the expected number of first generation offspring with magnitude $M \geq 4.35$ of a parent with magnitude $M<4.35$ is 0.008 (which corresponds to about 27 events in total). However, the i.i.d. magnitude distribution of the ETAS model allows these larger events to still be triggered by smaller earthquakes potentially explaining the difference in the information gain for this magnitude class.

Overall, predicting more events is favourable for both of the considered data sets. Although we do not know the confidence intervals for the IGPT, due to the massive difference between the IGPT of the ETAS and MDFHP models, for every magnitude class, we can conclude the MDFHP has the superior retrospective predictive performance.

\section{Conclusion and Future Developments}

We have presented strong evidence, ranging from information criteria, residual analysis, and retrospective predictive performance, that the MDFHP is able to capture more features of two seismic data sets, with multiple mainshock aftershock sequences, than the ETAS model can. The reason appears to be the increased flexibility allowing events of varying sizes to trigger offspring in a magnitude dependent manner. The ability to detect and study how events of varying magnitudes influence the time varying behaviour of the process, and the fact that the magnitude distribution heavily depends on the history, are novel and important advantages of the MDFHP over what currently exists in the literature. Due to the fact it can easily describe the decrease in seismic activity seen in the Middle America Trench data set when the ETAS model cannot, we hypothesise that the multidimensional fractional Hawkes process is able to account for problems in the data set such as a poorly selected geographic study region, incompleteness in the catalogue or other anomalous time varying behaviour.

A major topic of future investigation, so that the MDFHP can be implemented more effectively, is to develop and compare methods to objectively select both the number of subprocesses and the magnitude range of each. We present several methods that could potentially be used to identify the magnitude ranges. The first method would be to select the subprocesses so that each has the same number of events, which is an easily implemented method for any $N_b \in \N^+$. An alternate method is to use maximum likelihood estimation on the mark distribution. We saw that the mark distribution was influential in determining which model had the best AIC and BIC. Implementation of this method may be challenging since it is not clear how the event time distribution will change or if the mark densities in the log-likelihood will be the dominant term as $N_b$ increases. Following optimal subprocess selection, forecasting will be the focus of future research and because of the superior retrospective predictive performance of the MDFHP we believe its forecasting ability will be good.    

Another possible avenue of future research is to construct a continuous mark space model as a limiting process. Both the Mittag-Leffler function, as well as the univariate and multivariate Hawkes processes, are well behaved analytically. Quantities of interest for the fractional Hawkes process can be derived (an example being the mean intensity \citep[e.g.][]{
chen2021fractional}) and the behaviour of Hawkes processes has been characterised asymptotically by functional central limit theorems \citep[e.g.][]{bacry2013some,gao2018functional}. Using and developing results such as these could allow one to evaluate the following limit. For a finite $N_b$ consider the superposition process governed by the intensity

\begin{multline}
   \tilde \lambda_{N_b}(t,M|\HH_t)= \sum_{i=1}^{N_b}f_i(M|\HH_t)\lambda_i(t|\HH_t)=\\
   \sum_{i=1}^{N_b}f_i(M|\HH_t)\lambda_{0i}+\sum_{i=1}^{N_b}\sum_{j=1}^{N_b}f_i(M|\HH_t)\alpha_{ij}\sum_{\ell:t_{\ell,j}<t}\exp\{\gamma_{ij}(M_{\ell,j}-M_0)\}c_{ij}f_{\beta_{ij}}(c_{ij}(t-t_{\ell,j}))
\end{multline}
where $f_i(M|\HH_t)$ is the magnitude density of the $i^{th}$ subprocess, which is not necessarily the same density as Equation \eqref{eq:markdensity}. The continuous mark space model is 
\begin{equation}\label{eq:MDFHPlim}
    \lambda(t,M|\HH_t)=\lim_{N_b \to \infty} \tilde \lambda_{N_b}(t,M|\HH_t).
\end{equation}
Assuming the limit exists, we expect this model to yield interesting magnitude dependent behaviour because interactions between different events would depend continuously on their magnitudes.

\subsection{Acknowledgements}

The authors wish to acknowledge the use of New Zealand eScience Infrastructure (NeSI) high performance computing facilities, consulting support and/or training services as part of this research.  We wish to especially acknowledge Alexander Pletzer, Callum Walley and Murray Cadzow, without whom parts of this research could not have happened. New Zealand's national facilities are provided by NeSI and funded jointly by NeSI's collaborator institutions and through the Ministry of Business, Innovation \& Employment's Research Infrastructure programme \url{https://www.nesi.org.nz}.

 Furthermore, the authors are also grateful to Mark Bebbington for his helpful comments.

The authors report there are no competing interests to declare.

\subsection{Data and Code Availability}
 
The data is all accessible from public repositories being \url{https://earthquake.usgs.gov/earthquakes/search/}  for the Japan and Middle America Trench data sets. 

The workflow for this study, including code, data and output files, is available in the GitHub repository \url{https://github.com/davlo199/MDFHPCode}.

\appendix
\clearpage
\section{Confidence Intervals}

\begin{table}[ht!]

        \centering\small
        \begin{tabular}{|ccccc|}
        \hline
         & \multicolumn{4}{c|}{Index} \\
          Parameter & 11 & 12 & 21 & 22 \\ 
           & & & &  \\
          \multicolumn{5}{|c|}{MDFHP Japan}\\
          $\hat\lambda_0$ & [0.014,0.059] & NA & NA & [0.077,0.124]\\ 
          $\hat\alpha$  & [0.003,0.017] & [0.053,0.221] & [0.002,0.012] &  [0.338,0.611] \\ 
          $\hat\gamma$ &  [1.828,2.509] & [0.001,17.82] & [2.494,3.060] & [0.379,2.426] \\ 
          $\hat \beta $ & [0.572,0.881] & [0.269,0.597] &  [0.696,0.950] & [0.454,0.607] \\ 
          $\hat c$ & [2.713,10.958] & [0.001,2.109] & [1.692,3.686] & [0.087,0.266] \\ 
          $\hat B$ & [2.517,2.755] & NA & NA & [2.560,2.772]\\
           & & & & \\
  
          \multicolumn{5}{|c|}{MDFHP Middle America Trench}\\
          $\hat\lambda_0$ & [0.066,0.096] & NA & NA & [0.025,0.094]\\ 
          $\hat\alpha$  & [0.027,0.060] & [0.025,0.070] & [0.083,0.159] &  [0.673,0.895] \\ 
          $\hat\gamma$ &  [1.152,1.542] & [1.707,7.521] & [1.062,1.372] & [0.018,8.711] \\ 
          $\hat \beta $ & [0.575,0.827] & [0.492,0.808] &  [0.597,0.765] & [0.582,0.663] \\ 
          $\hat c$ & [5.908,22.263] & [0.119,1.788] & [1.455,4.584] & [0.047,0.090] \\ 
          $\hat B$ & [2.412,2.526] & NA & NA & [7.796,7.881]\\
          \hline
        \end{tabular}
     \caption{ \small 90\% asymptotic confidence intervals for the maximum likelihood estimates of the MDFHP models fitted to the Japan and Middle America Trench data sets.} 
     \label{tab:MDFHPCI}
\end{table}

\section{Subprocess Comparison}\label{app:subprocess}

In this appendix we briefly compare multidimensional fractional Hawkes processes with differently selected subprocesses by examining the AIC, BIC, residual processes and information gain. In addition to the models already studied, for each data set we study two additional models with differently selected subprocesses (although we still only use two subprocesses to keep the parameter space relatively low dimensional). Specifically, if the subprocesses are discretised as $[M_0,M_1'),[M_1',10)$, for the Japan data set we use $M_1'=5.5,5.75,6$ and for the Middle America Trench $M_1'=4.35,4.55,4.75$. Furthermore, we refer to the different multidimensional fractional Hawkes process models by MDFHP$M_1'$. In which case the MDFHP5.5 and MDFHP4.35 are the MDFHP models studied in the main body of the text fitted to the Japan and Middle America Trench data sets respectively.

We first compare the AIC and BIC values of every model. The AIC and BIC values in Table \ref{tab:AICBICMDFHPapp} suggest that for the Japan data set the MDFHP5.5 performs best. Furthermore, all three MDFHP models outperform the ETAS model. For the Middle America Trench data set all three MDFHP models outperform the ETAS model and the MDFHP4.35 is the best performing model. 

\begin{table}[h!]
        \centering\small
        \begin{tabular}{|cccc|}
        \hline
          Catalogue &Model &  AIC & BIC \\
          & &   & \\
          \multirow{4}*{Japan} &ETAS    & 3229.0 & 3260.9\\ 
          &MDFHP5.5  & 3146.0 & 3252.3\\ 
         
          &MDFHP5.75 & 3151.6 & 3257.8\\
          &MDFHP6 & 3151.6 & 3257.8 \\ 
          & &   & \\
         \multirow{4}*{Middle America Trench}& ETAS  & 5784.7 & 5822.7 \\
          &MDFHP4.35 & 4967.8 & 5094.4\\
           &MDFHP4.55 & 5070.5 & 5197.0 \\
            &MDFHP4.75 & 5107.3 & 5233.8 \\\hline
        \end{tabular}
        \caption{\small AIC and BIC scores for the ETAS and MDFHP models.}
        \label{tab:AICBICMDFHPapp}
\end{table}

\begin{figure}
    \centering
    \includegraphics[width=0.9\textwidth]{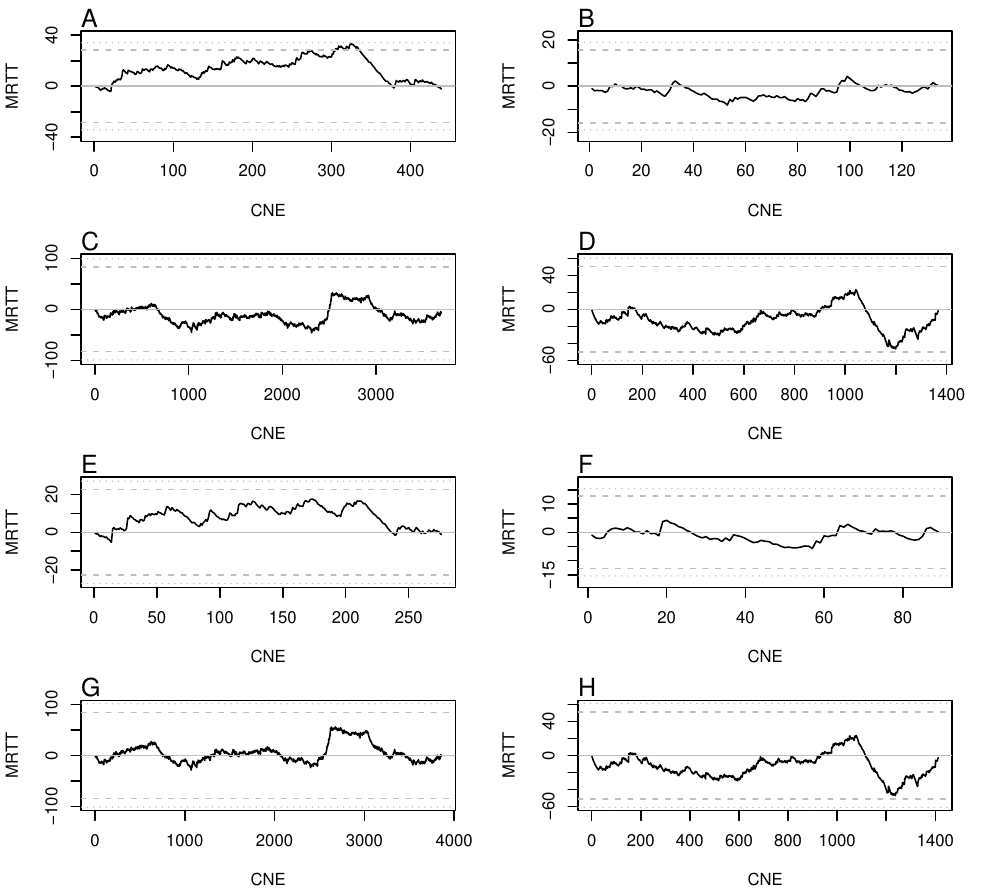}
    \caption{\small Mean removed transformed time (MRTT) are the black curves against cumulative number of events (CNE). The solid grey line is the theoretical mean, the dashed and dotted grey lines are the 95\% and 99\% confidence intervals respectively. (A,C,E,G) correspond to the Middle America Trench and (B,D,F,H) correspond to the Japan data set. For the following SP1 and SP2 stand for subprocess 1 and 2 respectively. (A) MDFHP4.55 SP1, (C) MDFHP4.55 SP2, (E) MDFHP4.75 SP1, (G) MDFHP4.75 SP2. (B) MDFHP5.75 SP1, (D) MDFHP5.75 SP2, (F) MDFHP6 SP1, (H) MDFHP6 SP2. }
    \label{fig:appresid}
\end{figure}

We now conduct residual analysis for the four additional models by using the same three residual tests as in Subsection \ref{subsec:ch6ResidAnalysis}. We include the graphical test in Figure \ref{fig:appresid} as well as the test statistics and $p$ values for the quantitative hypothesis tests in Table \ref{tab:PearsonPvaluesMDFHPApp}. At the 99\% confidence level, Figure \ref{fig:appresid} qualitatively suggests all subprocesses are describing the main features of the data sufficiently well. We now turn to the quantitative tests and find that at a 95\% confidence level we do not have sufficient evidence to reject the null hypothesis that the transformed uniform inter-event times of every subprocess are uniform. However, we have evidence to reject the null hypothesis that $U_k$ and $U_{k+1}$ are independent for subprocess 2 for both of the MDFHP5.75 and MDFHP6 models. Therefore, the main features of the Japan data set do not appear to be described sufficiently well by these two additional MDFHP models. For the Middle America Trench data set at the 99\% confidence level we do not have sufficient evidence to reject any of the null hypotheses, suggesting that both of these MDFHP models are able to describe the data sufficiently well.

\begin{table}

\centering\small
\begin{tabular}{|cccccc|}
\hline
Catalogue&Model & \multicolumn{2}{c}{Pearson Correlation} & \multicolumn{2}{c|}{KS test}\\
& & Test statistic& $p$ value & Test statistic & $p$ value\\

\multirow{4}*{Japan} &MDFHP5.75 SP1 & -0.63&0.53 & 0.05 & 0.91 \\
&MDFHP5.75 SP2 & 2.17& 0.03 & 0.028 & 0.25\\
&MDFHP6 SP1 & -1.21 & 0.23 & 0.06 & 0.89 \\
&MDFHP6 SP2 & 2.66 & 0.008 & 0.026 & 0.30\\
& & & & & \\
\multirow{4}*{Middle America Trench}& MDFHP4.55 SP1 & 1.14 & 0.26 & 0.03 & 0.87 \\
& MDFHP4.55 SP2 & -0.04 & 0.96 & 0.008 & 0.93 \\
& MDFHP4.75 SP1 & 1.651 & 0.100 &0.029 & 0.972\\
& MDFHP4.75 SP2 & 0.301& 0.763&0.013 &0.515\\
\hline
\end{tabular}
\caption{\small Test statistics and $p$ values of the Pearson test for serial correlation between $U_k$ and $U_{k+1}$ and KS test for uniformity of $U_k$ where SP1 and SP2 stand for subprocess 1 and subprocess 2 respectively.}
\label{tab:PearsonPvaluesMDFHPApp}
\end{table}

We finally present the IGPT for each of the three magnitude classes considered in Section \ref{subsec:ch6predperformance} for all three MDFHP models fit to both data sets in Table \ref{tab:mdfhp55J93IGPAppendix}. Within error tolerances there is likely no best predictive MDFHP model, and if there is it is not vastly superior to the other considered models. Overall, the information criteria and residual analysis would suggest it is more favourable for $M'_1$ to be smaller which allows more interaction between the subprocesses. Furthermore, as a rule of thumb it would appear that if, for a two subprocess model, one subprocess has less than 20\% of the data the model's performance suffers.

\begin{table}
    \centering\small
    \begin{tabular}{|cccccccc|}
    \hline
     Mag. class & $N_S(k)$ & $G_S(k)$ &$N_F(k)$ & $G_F(k)$ & $N(k)$ & $G(k)$ & $\rho_T(k)$ \\
      & & & & & & & \\
    \multicolumn{8}{|c|}{MDFHP5.5 (Japan)}\\

       $4.75 \leq M < 5.5$ & 631 & -159.18 & 1045 & 253.30 & 1676 & 94.12 & 0.028\\
        $5.5 \leq M < 6.5$ & 183 & 66.00 & 1493 & -45.13& 1676 & 20.86 & 0.0062\\
        $6.5 \leq M$ &  23 & 8.32 & 1653 & -2.52 & 1676 & 5.79 & 0.0017 \\
        & & & & & & & \\
    
        \multicolumn{8}{|c|}{MDFHP5.75 (Japan)}\\

        $4.75 \leq M < 5.5$ & 631 & -144.7 & 1045 & 239.49 & 1676 & 94.78 & 0.028\\
        $5.5 \leq M < 6.5$ & 183 & 48.14 & 1493 & -27.38& 1676 & 20.77 & 0.0062\\
        $6.5 \leq M$ &  23 & 12.67 & 1653 & -5.54 & 1676 & 7.13 & 0.0021 \\
         & & & & & & & \\
\multicolumn{8}{|c|}{MDFHP6 (Japan)}\\

        $4.75 \leq M < 5.5$ & 631 & -144.61 & 1045 & 240.59 & 1676 & 95.98 & 0.029\\
        $5.5 \leq M < 6.5$ & 183 & 46.90 & 1493 & -28.39& 1676 & 18.52 & 0.0055\\
        $6.5 \leq M$ &  23 & 8.52 & 1653 & -2.99 & 1676 & 5.53 & 0.0017 \\
         & & & & & & & \\
 \multicolumn{8}{|c|}{MDFHP4.35 (Middle America Trench)}\\

        $4 \leq M < 4.35$ & 1705 & 27.49 & 1296 & 124.15 & 3001 & 151.64 & 0.025\\
        $4.35 \leq M < 5.35$ & 569 & 106.86 & 2432 & -90.00& 3001 & 16.86 & 0.003\\
        $5.35 \leq M$ &  71 & 157.53 & 2930 & -56.78 & 3001 & 100.75 & 0.017 \\

 & & & & & & & \\
    
     \multicolumn{8}{|c|}{MDFHP4.55 (Middle America Trench)}\\

        $4 \leq M < 4.35$ & 1705 & 64.13 & 1296 & 86.17  & 3001 & 150.30 & 0.025\\
        $4.35 \leq M < 5.35$ & 569 & 34.70 & 2432 &-20.44 & 3001 & 14.25 & 0.002 \\
        $5.35 \leq M$ &  71 & 171.37 & 2930 & -67.74 & 3001 & 103.63 &  0.017\\

 & & & & & & & \\

      \multicolumn{8}{|c|}{MDFHP4.75 (Middle America Trench)}\\

        $4 \leq M < 4.35$ & 1705 & 63.68 & 1296 & 85.97 & 3001 & 149.66 & 0.025\\
        $4.35 \leq M < 5.35$ & 569 & 36.47 & 2432 &-25.70 & 3001 & 10.77 & 0.002\\
        $5.35 \leq M$ &  71 & 169.67 & 2930 &-68.38  & 3001 & 101.29 & 0.017 \\
 \hline
\end{tabular}
    \caption{\small Information gain of the three magnitude classes for all considered models. For the $k^{th}$ magnitude class $N_S(k)$ is the number of intervals such that $X_i(k)=1$ and $G_S(k)$ is the gain in these intervals. $N_F(k)$ is the number of intervals such that $X_i(k)=0$ and $G_F(k)$ is the gain in these intervals. $N(k)$ is the total number of intervals, $G(k)$ is the total information gain, and $\rho_T(k)$ is the information gain per unit time.}
    \label{tab:mdfhp55J93IGPAppendix}
\end{table}
\clearpage

\bibliography{Bibliography-MM-MC}

\end{document}